\documentclass[floats,twocolumn,prb]{revtex4}
\usepackage{graphicx}
\usepackage{dcolumn}
\usepackage{bm}
\usepackage{mathrsfs}

\newcommand{\mycite}[1]{\scalebox{1.4}[1.4]{\raisebox{-0.85ex}{\cite{#1}}}}

\begin{document}
\preprint{APS/123-QED}

\title{Localization and Orbital Selectivity in Iron-Based Superconductors
with Cu Substitution}

\author{Yang Liu$^1$}
\author{Da-Yong Liu$^2$}
\author{Jiang-Long Wang$^3$}
\author{Jian Sun$^{1}$}
\author{Yun Song$^{1}$}
\thanks{yunsong@bnu.edu.cn}
\author{Liang-Jian Zou$^2$}

\affiliation{%
$^1$Department of Physics, Beijing Normal University, Beijing 100875, China}%

\affiliation{%
$^2$Key Laboratory of Materials Physics, Institute of Solid State Physics,
 Chinese Academy of Sciences, P. O. Box 1129, Hefei 230031, China}%

\affiliation{%
$^3$College of Physics Science and Technology, Hebei University, Baoding 071002,
China}%

\date{\today}

\begin{abstract}
We study an inhomogeneous three-orbital Hubbard model for the
Cu-substituted iron pnictides using an extended real-space Green's
function method combined with density functional calculations. We
find that the onsite interactions of the Cu ions are the principal
determinant of whether an electron dopant or a hole dopant is caused
by the Cu substitution. It is found that the Cu substitution could
lead to a hole doping when its onsite interactions are smaller than
a critical value, as opposed to an electron doping when the
interactions of Cu ions are larger than the critical value,
which may explain why the effects of Cu substitution on the carrier
density are entirely different in NaFe$_{1-x}$Cu$_x$As and
Ba(Fe$_{1-x}$Cu$_x$)$_2$As$_2$. We also find that the effect of a
doping-induced disorder is considerable in the Cu-substituted iron
pnictides, and its cooperative effect with electron correlations
contributes to the orbital-selective insulating phases in
NaFe$_{1-x}$Cu$_x$As and Ba(Fe$_{1-x}$Cu$_x$)$_2$As$_2$.
\end{abstract}


\maketitle


\section{INTRODUCTION}
\label{sec:INTR}

Iron-based superconductors \cite{Hosono1,Hosono2,ChenX} have
attracted great attention in recent years due to their considerable
high superconducting transition temperatures and their rich phase
diagrams, which are qualitatively similar to that of cuprate
superconductors.\cite{Chu,Hirschfeld} In cuprate superconductors,
substituting Cu with other transition metals is an effective way to
gain a better insight into the origin of the high-temperature
superconductivity.\cite{Alloul,Balatsky} Likewise, many experiments
were conducted to reveal the effects of substituting Fe with other
transition metals in some families of iron-based superconductors.
\cite{McLeod,YanYJ,ChengP,Merz,Ideta,WangAF,HuangTW,NiN,LiJ,CuiST,Williams}
Among various transition-metal substitutions, Cu doping is highly
disruptive to the electron structures of the FeAs sheets. There is
lively debate on whether electron doping or hole doping occurs in
Cu-substituted Fe-based superconductors.
\cite{McLeod,YanYJ,ChengP,Ideta,Merz}

On the one hand, the X-ray photoelectron spectra\cite{McLeod,YanYJ}
show that Cu has a closed $3d^{10}$ shell in
Ba(Fe$_{1-x}$Cu$_x$)$_2$As$_2$ and SrFe$_{2-x}$Cu$_x$As$_2$
compounds, suggesting that Cu substitution behaves like hole doping.
On the other hand, in the 122 phase, both Cu and Ni substitutions
introduced a significant electron doping effect in
Ba$_{0.6}$K$_{0.4}$Fe$_2$As$_2$ and suppressed the superconducting
transition temperature considerably. \cite{ChengP} Nevertheless, the
angle resolved photoemission spectroscopy (ARPES) experiment on
Ba(Fe$_{1-x}$Cu$_x$)$_2$As$_2$ predicted that a part of the
electrons in Cu substitution preferentially occupied the Cu $3d$
states and did not behave like a mobile carrier.\cite{Ideta,Merz}
Moreover, the measurements on a series of NaFe$_{1-x}$Cu$_x$As
single crystals showed that Cu substitution serves as an effective
electron dopant with strong impurity potential.\cite{WangAF} Such an
observation was also supported by the ARPES study, which exhibited
the increasing of the Fermi level in NaFe$_{1-x}$Cu$_x$As,
indicating a fraction of electron doping introduced by the Cu
dopant. \cite{CuiST}  Up to date, the dispute on the electron or
hole doping in the 111 and 122 phases is far from resolved.
Therefore, further investigations are needed to explain the reason
why the Cu substitution plays different roles in
NaFe$_{1-x}$Cu$_x$As and Ba(Fe$_{1-x}$Cu$_x$)$_2$As$_2$ compounds.

One also notices that the random distribution of Cu over Fe has been
confirmed by the X-ray absorption near-edge structure in the Fe $K$
edge in Fe$_{1-x}$Cu$_x$Se$_{1-\delta}$. \cite{HuangTW} The
first-principles calculation also highlighted the necessity of
including disorder effects on the iron-based superconductors with
transition-metal substitution. \cite{Berlinjn} Therefore, the
quantum phase transition introduced by substituting Fe with Cu is
complicated because Cu doping not only changes the carrier density
but also leads to strong disorder effect. Additionally, the
first-principles calculation demonstrated that the effective on-site
Coulomb interactions in some families of Fe-based superconductors
were considerably strong, \cite{Miyake} leading the parent compounds
to be very close to the Mott insulating phase. \cite{Nakajima} The
role of electronic correlation in the renormalizing of electronic
bandwidths, \cite{Werner} magnetic moments \cite{LiuDY},
orbital-selective Mott phase, \cite{Medici-2009,Medici-2014} and
orbital order \cite{QuanYM} in iron-pnictides is also recognized.
More interestingly, heavy Cu doping drives the occurrence of a
metal-insulator transition (MIT), \cite{WangAF} which rarely happens
in iron-pnictide compounds. This naturally raises a question as to
whether the substitution of Fe with Cu drives the system to transit
from a metal to an insulator, as was observed by Wang {\it et al.},
\cite{WangAF} and what the nature of the MIT is.

To understand the mysterious electron or hole doping and the
underlying physics of the MIT tuned by substituting Fe with Cu in an
Fe-based superconductor, we proposed an inhomogeneous three-orbital
Hubbard model after reconstructing the Fermi surfaces of NaFeAs and
NaCuAs based on the first-principles calculations. Here we develop
an extended real-space Green's function method, which is an
adequate approach for the study of the cooperative effect of the
multi-orbital electronic correlations and doping-induced disorder.
We find that the increasing of the onsite interactions $u$ of the
Cu ions can transfer electrons of Cu ions to Fe ions, and the
critical point of the inversion from hole doping to electron doping
is about $u_c$=5.2 eV when the interactions of Fe ions are
$U$=1.5 eV and $J=U/8$. It is found that the Cu substitutions lead to a
hole doping when $u<$5.2 eV, as opposed to an electron doping when
$u\geq$5.2 eV, where the carrier occupancy of the Cu sites drops to
zero. With the increasing
of Cu substitutions,  an enhancement of the Anderson localization in
$d_{xz}$ and $d_{yz}$ orbitals is observed, whereas a Mott gap is
found in the $d_{xy}$ orbital, indicating that the cooperative
effect of multiorbital correlations and doping-induced disorder can
lead to an orbital-selective insulating phase in the Cu-substituted
Fe-based superconductor.

This paper is organized as follows. In section~\ref{Sec:Model}, we
first present the electronic structures of compounds
NaFe$_{1-x}$Cu$_x$As obtained by the first-principles calculations;
we then propose an inhomogeneous three-orbital Hubbard model to
study the cooperative effects of the multi-orbital correlations and
doping-induced disorder in Cu-substituted iron pnictides. The
extended real-space Green's function approach employed in this paper
is introduced in section~\ref{Sec:RSGF}.  In section~\ref{Sec:U+W},
we study the effects of Cu substitution on the carrier density and
discuss the orbital-selective insulating phase introduced by the
cooperative effects of multi-orbital interactions and doping-induced
disorder. The principal findings of this paper are summarized in
section~\ref{Sec:Con}.


\section{LDA bands and model Hamiltonian}
\label{Sec:Model}

\begin{figure}[htbp]
\includegraphics[scale=0.3]{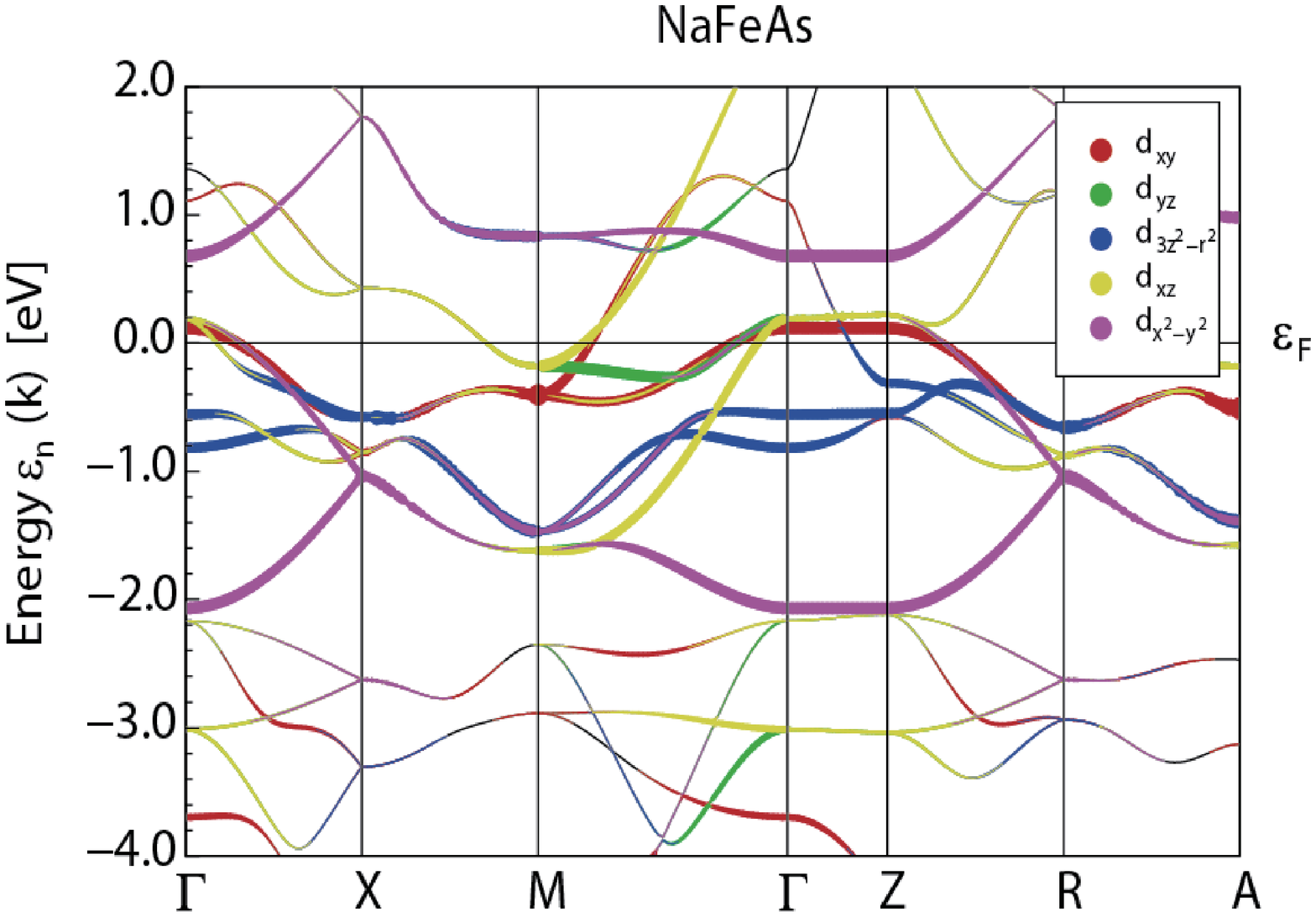}
\includegraphics[scale=0.3]{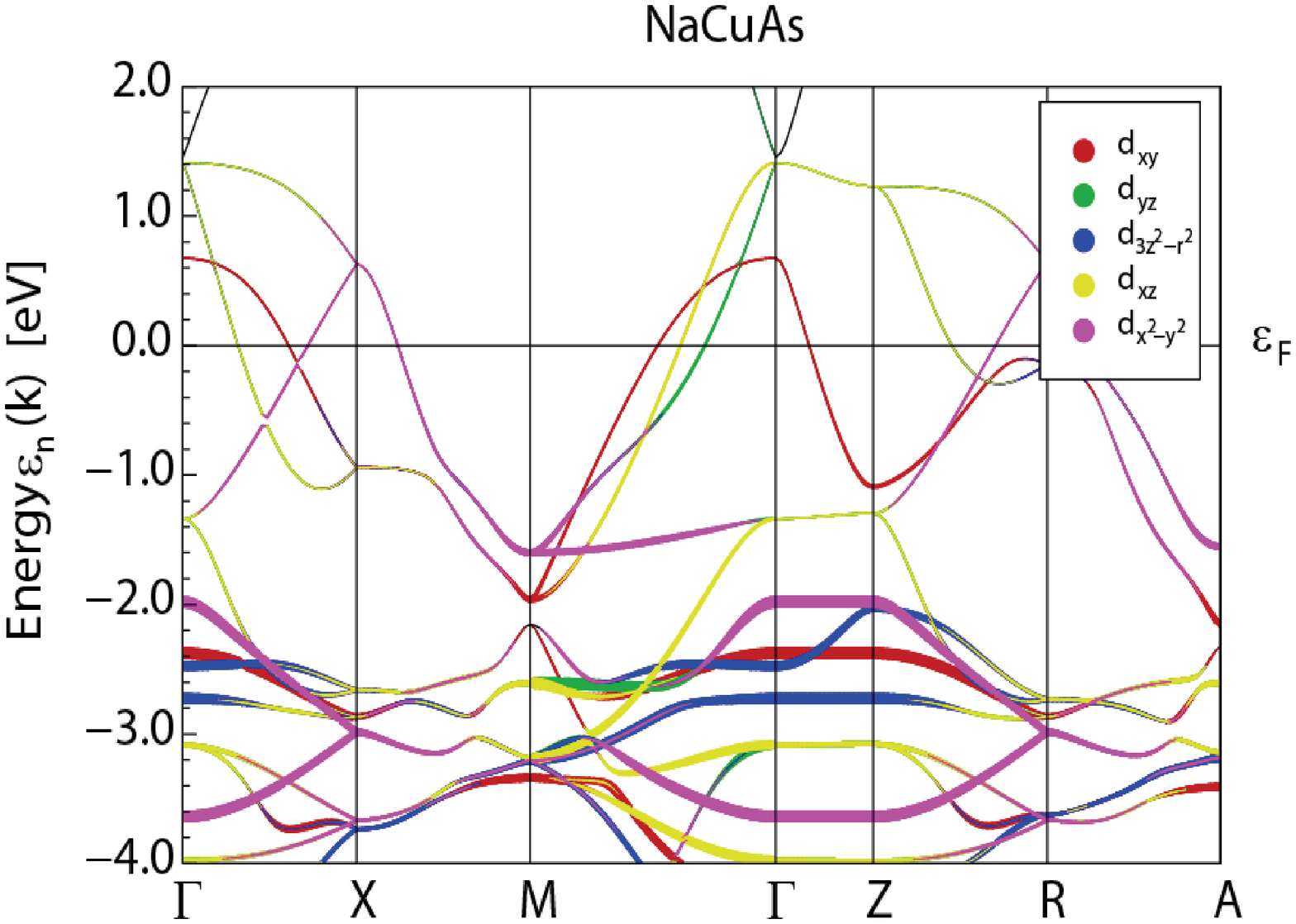}
\caption{(Color online) Band structures of NaFeAs (upper) and NaCuAs (lower)
obtained by the GGA approach. Thick lines project out the Cu $3d$ orbitals
in the lower figure.}
\label{fig:BS-FeCu}
\end{figure}

We first perform the local density functional calculation to
understand the primary electronic structures of Cu-doped compounds
NaFe$_{0.9}$Cu$_{0.1}$As,  NaFe$_{0.8}$Cu$_{0.2}$As,
NaFe$_{0.7}$Cu$_{0.3}$As, and NaCuAs, as well as the parent compound
NaFeAs. The band structures are calculated by the full potential
local orbital (FPLO9) code \cite{Koepernik-1999} using the Perdew-Wang 92
version \cite{Perdew} of the local-density approximation (LDA) for
the exchange and correlation potentials. The orbitals of sodium
(2$s$, 3$s$, 4$s$, 2$p$, 3$p$, 4$p$, and 3$d$), iron (3$s$, 4$s$,
5$s$, 3$p$, 4$p$, 5$p$, 3$d$, and 4$d$), and arsenic (3$s$, 4$s$,
5$s$, 3$p$, 4$p$, 5$p$, 3$d$, and 4$d$) ions are treated as valence
states, while the other lower-lying orbitals are considered as core
states. Utilizing the lattice constants and atomic positions
obtained by experiments, \cite{WangAF} the self-consistency process
counts a total of 10416 irreducible $k$-points for the $k$-space
integrations.

\begin{figure}[htbp]
\includegraphics[scale=0.45]{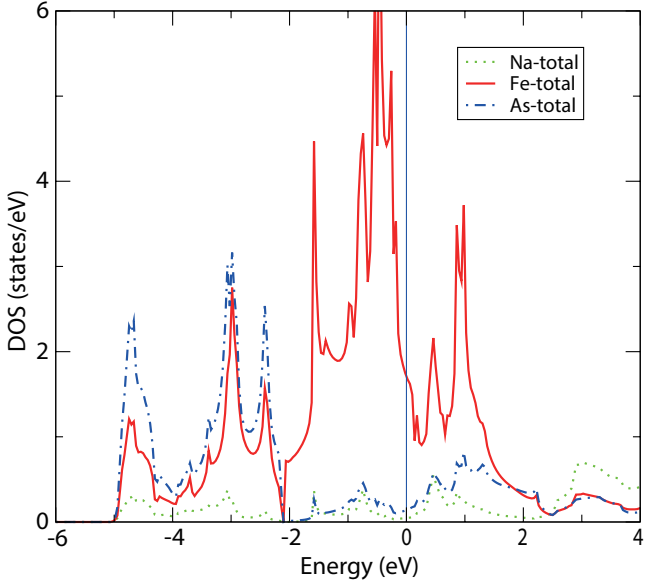}
\caption{(Color online) Density of states of NaFeAs obtained by the GGA approach.
Three strongly hybridized Fe $3d$ orbitals $d_{xy}$, $d_{xz}$, and $d_{yz}$
determine the Fermi surfaces of NaFeAs.}
\label{fig:DOS-NaFeAs}
\end{figure}

We plot the band structures of NaFeAs and NaCuAs
in the upper and lower panels of Fig.~\ref{fig:BS-FeCu}, respectively.
The heavy stripes represent the weights of Fe and Cu 3$d$ orbitals
contributed to the bands. The weights are normalized, that is, the
summation of the weights of all valence orbitals at a given point of
the band structure should be unity. From Fig.~\ref{fig:BS-FeCu}, we
find that the principal bands that determine the Fermi surfaces of NaFeAs
comprise three strongly hybridized Fe $3d$ orbitals $d_{xy}$,
$d_{xz}$, and $d_{yz}$.  Moreover, it is shown in Fig.~\ref{fig:BS-FeCu}
that the Cu 3$d$ orbitals primarily distribute
from -2 eV to -4 eV below the Fermi level, suggesting that 3$d$
orbitals of doped Cu ions sink below Fermi level.
In addition, we find that the weight of those bands around the Fermi level
comes from As 4$p$ orbitals. The Cu 4$s$ and Na 3$s$ orbitals are higher in
energy and irrelevant. We also plot the total density of states (DOS) of
NaFeAs in Fig.~\ref{fig:DOS-NaFeAs}.
It is obvious that the DOS of Fe 3$d$ orbitals at the Fermi energy is
dominant. While, the DOS of Na 3$s$ and As 4$p$ orbitals near the Fermi
level are both very small, indicating that these orbitals could be ignored.

The coherent potential approximation (CPA) \cite{Soven,Taylor} is
employed to account for the effect of substitutional disorder
introduced by the Cu doping at Fe sites in NaFe$_{1-x}$Cu$_{x}$As.
As implemented in the FPLO5 code, \cite{Koepernik-1997}
the CPA approach treats approximately the disordered crystal as an
effective medium. \cite{Soven,Taylor}
In the CPA calculation of the variation of electronic structure
caused by Cu doping, the lattice parameters of NaFe$_{1-x}$Cu$_{x}$As
($x$=0.1, 0.2, and 0.3) are fixed at the experimental values of the
parent compound NaFeAs, and 1496 irreducible $k$-points are counted.

\begin{figure}[htbp]
\includegraphics[scale=0.4]{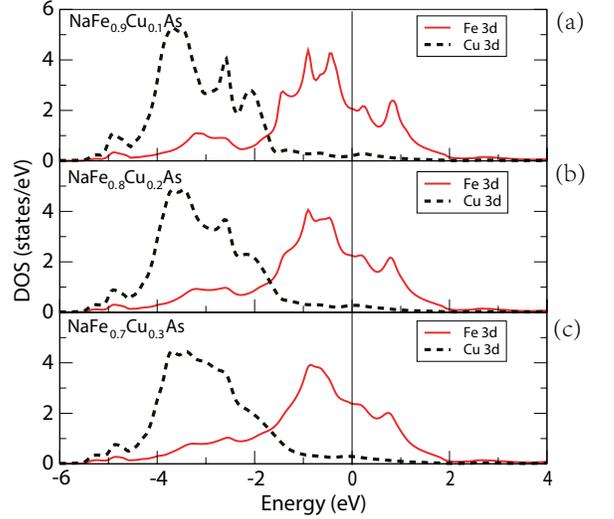}
\caption{(Color online) The density of states of compounds
NaFe$_{1-x}$Cu$_{x}$As at $x$=0.1 (a), $x$=0.2 (b), and $x$=0.3 (c) within the coherent
potential approximation.}
\label{fig:DOSLDA-Cu}
\end{figure}

In Fig.~\ref{fig:DOSLDA-Cu}, we show the DOS of
NaFe$_{1-x}$Cu$_{x}$As at $x$=0.1 and 0.3 near the Fermi level,
respectively. With the increasing of Cu-doping concentration, the
energy distribution of Cu 3$d$ orbitals does not vary significantly,
clearly indicating that the Cu 3$d$ orbitals always lie far below
E$_F$. We do not observe MIT over a wide region of Cu-doping
concentration, suggesting that the filling-factor controlled MIT
could be ruled out, and thus
the electronic correlation or disorder driven MIT is the most
relevant candidate.

The band structure and orbital composition obtained by the
first-principles calculation for undoped NaFeAs can be reproduced by
a three-orbital tight-binding (TB) model. The TB model parameters
are shown in Table~\ref{Tb:pmts}, which are very similar to the
model parameters of Ref.~[\mycite{Daghofer-2012}]. Based on our LDA
calculations of NaFe$_{1-x}$Cu$_x$As, we introduce an inhomogeneous
three-orbital Hubbard model for the Cu-substituted Fe-based
compounds, where a certain percentage of Fe sites replaced by Cu
ions are chosen at random. Therefore, the whole Hamiltonian consists
of three parts,
\begin{equation}
H=H_{\textrm{\tiny{Fe}}}+H_{\textrm{\tiny{Cu}}}+H_{\textmd{\tiny{Hyb}}},
\label{Eq_HamW}
\end{equation}
where $H_{\textrm{\tiny{Fe}}}$, $H_{\textrm{\tiny{Cu}}}$, and
$H_{\textmd{\tiny{Hyb}}}$ represent the Hamiltonians of Fe ions,
Cu-doping constitutions, and the hybridizations between Fe and Cu
sites, respectively.

The partial Hamiltonian
$H_{\textrm{\tiny{Fe}}}$ for the electrons in $d_{xy}$, $d_{xz}$,
and $d_{yz}$ orbitals of Fe ions can be expressed as
\begin{eqnarray}
H_{\textmd{\tiny{Fe}}}&=&-\sum_{i\neq j (i,j\in Fe)}
     \sum_{\alpha\beta\sigma}
     T_{i\alpha,j\beta}c_{i\alpha\sigma}^\dag c_{j\beta\sigma}\nonumber\\
  &&-\sum_{i\in Fe,\alpha\sigma}\mu_{\alpha}n_{i\alpha\sigma}
     +U\sum_{i\in Fe,\alpha}n_{i\alpha\uparrow}n_{i\alpha\downarrow}\nonumber\\
  &&+\sum_{i\in Fe}\sum_{\alpha<\beta,\sigma\sigma'}(U'-J \delta_{\sigma\sigma'})
    n_{i\alpha\sigma}n_{i\beta\sigma'},
 \label{Eq_HamFe}
\end{eqnarray}
where $c_{i\alpha\sigma}^\dag$ ($c_{i\alpha\sigma}$) creates
(annihilates) an electron with spin projection $\sigma$ for orbital
$\alpha$ ($d_{xz}$,$d_{yz}$, or $d_{xy}$) of an iron site $i$, and
$n_{i\alpha\sigma}=c_{i\alpha\sigma}^\dag c_{i\alpha\sigma}$
represents the corresponding electron number operator.
$T_{i\alpha,j\beta}$ represent the hopping integrals of electrons
between nearest-neighbor (NN) or next-nearest-neighbor (NNN) iron
sites $i$ and $j$ within orbitals $\alpha$ and $\beta$.
\cite{Daghofer-2010} $\Delta_{xy}=\mu_{xy}-\mu_{xz}$ is the energy
difference between the $d_{xy}$ and degenerate $d_{xz}$/$d_{yz}$
orbitals. More details regarding the definition of kinetic energy in
a Hamiltonian~(\ref{Eq_HamFe}) can be found in
appendix~\ref{APP-HT}. $U$ and $U^{\prime}$ are the on-site
intra-orbital and inter-orbital Coulomb interactions on iron sites,
respectively, and $J=J_z$ represents the Ising-type Hund's
couplings. In our calculations, we ignore spin-flip and pair-hopping
processes. \cite{Costi,Bouadim,HLee}

\begin{table}[htbp]
\caption{\label{Tb:pmts} Parameters for the hopping integrals of the
three-orbital Hubbard model. \cite{Daghofer-2012,Daghofer-2010}
The energy unit is electron volts. The quantities are defined in appendix~\ref{APP-HT}. }
\begin{tabular}{p{0.65cm}p{0.65cm}p{0.9cm}p{0.65cm}p{0.85cm}p{1.0cm}p{1.1cm}p{0.85cm}p{0.65cm}p{0.5cm}}
\hline \hline  \\ [-3pt]
 $  T_1$  & $T_2$ & $T_3$   &  $T_4$  & $T_5$   & $T_6$  & $T_7$   & $T_8$  &   $\Delta_{xy}$ \\
 \hline \\ [-3pt]
 $ 0.15$   & 0.15 & $-$0.12 &  $0.06$ & $-$0.08 & 0.1825 & 0.08375 & $-$0.03 &   0.75          \\
\hline\hline
\end{tabular}
\end{table}

As shown by the first-principles calculations, the hybridization is
mainly between the Fe $3d$ orbitals and the Cu $4s$ orbitals, but it
is very weak. The Cu $3d$ orbitals distribute from -2 eV to -4 eV
below the Fermi level, unlike the Cu ions in cuprate high-$T_c$
superconductors. When an Fe ion is substituted by a Cu ion in
Fe-based superconductors, we need to consider all three $t_{2g}$
orbitals of a Cu ion instead of the single $d_{x^2-y^2}$ orbital.
Therefore, the Hamiltonian for the Cu substitutions is
\begin{eqnarray}
H_{\textmd{\tiny{Cu}}}&=&-\sum_{i\neq j (i,j\in Cu)}
    \sum_{\gamma\sigma}
    t_{i j}d_{i\gamma\sigma}^\dag
    d_{j\gamma\sigma}\nonumber\\
  &&-\sum_{i\in Cu,\gamma\sigma} \nu_{\gamma}
    n_{i\gamma\sigma}^d
    +u\sum_{i\in Cu,\gamma} n_{i\gamma\uparrow}^d
    n_{i\gamma\downarrow}^d\nonumber\\
 &&+\sum_{i\in Cu}\sum_{\gamma<\lambda,\sigma\sigma'}(u'-j \delta_{\sigma\sigma'})
 n^d_{i\gamma\sigma}n^d_{i\lambda\sigma'},
 \label{Eq_HamCu}
\end{eqnarray}
where $d_{i^\prime\gamma\sigma}^\dag$ ($d_{i^\prime\gamma\sigma}$)
creates (annihilates) an electron at a copper site $i^\prime$.
$\nu_{\gamma}$ are the chemical potentials of different orbitals
$\gamma$. The intraorbital hopping integral between two copper sites
is indicated as $t_{i^\prime j^\prime}$, and the intraorbital and
interorbital interactions, as well as the Hund's rule couplings, are
represented by $u$, $u^{\prime}$, and $j$ for copper sites,
respectively.

As our calculations show that the intraorbital
interactions $u$ play a principal role, only the effect of $u$ is
taken into account. That is, we choose $u^{\prime}=u$ and $j=0$ in our
discussion of Cu substitution effect in the iron pnictides. Because
the Cu $3d$ orbitals locate about 3 eV below the Fermi level, the
intraorbital hopping integrals $t_{ij}$ are predicted to be quite small.
Owing to the restriction of the precision limitation of our
calculations for the disordered cases, we approximate $t_{ij}$ in the
range of 0 to 0.03 eV for the NN hopping integrals between Cu ions.
In our calculations, we
choose $t$=0.01 eV for the NN hopping terms but $t=0$ for other
long-distance hopping terms. We find that the value of
$t$ only has an extremely-limited effect on the bandwidth of the Cu
impurity band, and its effect on the DOS of Fe is negligible when
the concentration of Cu substitution is less than 0.2.

Moreover, the hybridization of Fe and Cu ions
can be expressed as
\begin{eqnarray}
H_{\textmd{\tiny{Hyb}}}=- \sum_{i\in Fe,\alpha}\sum_{j\in Cu,\gamma\sigma}
  t_{i j}^{\prime} (c_{i\alpha\sigma}^\dag
  d_{j\gamma\sigma}+\textmd{H.c.}).
 \label{Eq_Hamex}
\end{eqnarray}
Because of the very weak effects of the long-distance terms, we only
need to consider the NN hopping integrals $t^{\prime}$ between Fe
and Cu ions, which are also assumed as 0.01 eV. Just the same as $t$,
the effect of $t^{\prime}$ on the DOS of Fe ions is also negligible
for the case with $x<0.2$. We study the
inhomogeneous three-orbital Hubbard model Eq.(\ref{Eq_HamW}) by
developing a real-space Green's function method, which is described
in the next section.


\section{Real-Space Green's Function Approach}
\label{Sec:RSGF}

For a three-orbital Hubbard model, the real-space Green's function
of a square lattice with $N=L^2$ sites is expressed as a $3N\times
3N$ matrix
\begin{widetext}
\begin{equation}
\mathscr{G}=
\left(\begin{array}{ccccccc}
 G_{11}     &\cdots    &G_{1,h-1}    &F^*_{1h}    &G_{1,h+1}
&\cdots   &G_{1M}       \\
\vdots     &\ddots    &\vdots        &\vdots      &\vdots
&\ddots   &\vdots       \\
G_{h-1,1}  &\cdots    &G_{h-1,h-1}   &F^*_{h-1,h} &G_{h-1,h+1}
&\cdots   &G_{h-1,M}    \\
F_{h1}     &\cdots    &F_{h,h-1}     &D_{hh}      &F_{h,h+1}
&\cdots   &F_{hM}       \\
G_{h+1,1}  &\cdots    &G_{h+1,h-1}   &F^*_{h+1,h} &G_{h+1,h+1}
&\cdots   &G_{h+1,M}    \\
\vdots     &\ddots    &\vdots        &\vdots      &\vdots
&\ddots   &\vdots       \\
G_{M1}     &\cdots    &G_{M,h-1}     &F^*_{Mh}    &G_{M,h+1} &\cdots
&G_{MM}
\end{array}\right)_{M=3N},
\label{G}
\end{equation}
\end{widetext}
where $h$ represents the sites occupied by Cu ions. $G_{ij}$, $F_{ij}$, and
$D_{ij}$ are $3\times 3$ matrices with elements defined as
\begin{eqnarray}
G_{ij}^{\alpha\beta}&=&\langle\langle c_{i\alpha\sigma} \mid
c^{\dag}_{j\beta\sigma}\rangle\rangle\nonumber\\
F_{ij}^{\alpha\beta}&=&\langle\langle d_{i\alpha\sigma} \mid
c^{\dag}_{j\beta\sigma}\rangle\rangle\nonumber\\
D_{ij}^{\alpha\beta}&=&\langle\langle d_{i\alpha\sigma} \mid
d^{\dag}_{j\beta\sigma}\rangle\rangle .
\end{eqnarray}
Each element of the Green's function matrices is obtained by the
equation of motion, \cite{Zubarev} for example,
\begin{eqnarray}
(\omega &+&\mu_\alpha) \langle\langle c_{i\alpha\sigma}
 \mid c_{j\beta\sigma}^\dag\rangle\rangle
  =\delta_{ij}\delta_{\alpha\beta}
  -\sum_{b\in Cu,\gamma}t_{ib}'\langle\langle d_{b\gamma\sigma}
  \mid c_{j\beta\sigma}^\dag\rangle\rangle
 \nonumber\\
  && -\sum_{b\in Fe,m}T_{ib\alpha m}\langle\langle
  c_{bm\sigma}\mid c_{j\beta\sigma}^\dag\rangle\rangle
 \nonumber\\
  &&+U\langle\langle n_{i\alpha\bar{\sigma}} c_{i\alpha\sigma}
  \mid c_{j\beta\sigma}^\dag\rangle\rangle
  +U'\sum_{l\neq \alpha}\langle\langle
  n_{il\bar{\sigma}}c_{i\alpha\sigma}\mid
  c_{j\beta\sigma}^\dag\rangle\rangle
 \nonumber\\
   &&+(U'-J)\sum_{l\neq \alpha}\langle\langle n_{il\sigma}
   c_{i\alpha\sigma}\mid c_{j\beta\sigma}^\dag\rangle\rangle,
\label{Eq:FOEOM}
\end{eqnarray}
from which come the second-order Green's functions, i.e.
$\langle\langle n_{il\bar{\sigma}} c_{i\alpha\sigma}\mid
c_{j\beta\sigma}^\dag\rangle\rangle$ and $\langle\langle
n_{il\sigma} c_{i\alpha\sigma}\mid
c_{j\beta\sigma}^\dag\rangle\rangle$ on the right side of
Eq.~(\ref{Eq:FOEOM}). Similarly, the second-order Green's functions
can be obtained by their equations of motion, where we will find the
third-order Green's functions. Having an analogy with the Hubbard-I
approximation, \cite{Hubbard-I} we introduce a proper decoupling
process to make up a self-consistent loop for the calculations of
inhomogeneous Green's function $\hat{G}$. More details can be found
in Appendix~\ref{APP-DA}.


We thus obtain a complete and solvable set of equations for all
elements of the Green's function matrix $\hat{G}$ in Eq.(~\ref{G}),
\begin{widetext}
\begin{eqnarray}
M_\alpha \langle\langle c_{i\alpha\sigma}\mid
c_{j\beta\sigma}^\dag\rangle\rangle
&=&N_\alpha\delta_{ij}\delta_{\alpha\beta}
-K_\alpha\delta_{ij}\delta_{l\beta} -(N_\alpha-K_\alpha)(\sum_{b\in
Fe,m}T_{ib\alpha m}\langle\langle c_{bm\sigma}\mid
c_{j\beta\sigma}^\dag\rangle\rangle +\sum_{b\in
Cu,\gamma}t_{ib}'\langle\langle d_{b\gamma\sigma}\mid
c_{j\beta\sigma}^\dag\rangle\rangle)\nonumber\\
&&+\frac{U'-J}{A_\alpha}\sum_{l\neq\alpha} (-\sum_{b\in
Fe,m}T_{iblm}\langle  c_{bm\sigma}^\dag c_{i\alpha\sigma}\rangle
+\sum_{b\in Cu,\gamma}t_{ib}'\langle d_{b\gamma\sigma}^\dag
c_{i\alpha\sigma}\rangle)\langle\langle
c_{il\sigma}\mid c_{j\beta\sigma}^\dag\rangle\rangle,\nonumber\\
M_\alpha \langle\langle c_{i\alpha\sigma}\mid
d_{j\lambda\sigma}^\dag\rangle\rangle &=&-(N_\alpha-K_\alpha)(\sum_{b\in
Fe,m}T_{ib\alpha m}\langle\langle c_{bm\sigma}\mid
d_{j\lambda\sigma}^\dag\rangle\rangle
+\sum_{b\in Cu,\gamma}t_{ib}'\langle\langle  d_{b\gamma\sigma}\mid
d_{j\lambda\sigma}^\dag\rangle\rangle)\nonumber\\
&&+\frac{U'-J}{A_\alpha}\sum_{l\neq\alpha}(-\sum_{b\in Fe,m}
T_{iblm}\langle c_{bm\sigma}^\dag c_{i\alpha\sigma}\rangle
+\sum_{b\in Cu,\gamma}t_{ib}'\langle d_{b\gamma\sigma}^\dag
c_{i\alpha\sigma}\rangle)\langle\langle
c_{il\sigma}\mid d_{j\lambda\sigma}^\dag\rangle\rangle,\nonumber\\
(\omega+\nu_{\gamma})\langle\langle d_{i\gamma\sigma}\mid
c_{j\beta\sigma}^\dag\rangle\rangle &=&-(1+\frac{u\langle
n^d_{i\bar{\sigma}}\rangle}{\omega+\nu_{\gamma}-u}) (\sum_{b\in
Cu}t_{ib}\langle\langle d_{b\gamma\sigma}\mid
c_{j\beta\sigma}^\dag\rangle\rangle +\sum_{a\in
Fe,\alpha}t_{ai}'\langle\langle
c_{a\alpha\sigma}\mid c_{j\beta\sigma}^\dag\rangle\rangle),\nonumber\\
(\omega+\nu_{\gamma})\langle\langle d_{i\gamma\sigma}\mid
d_{j\lambda\sigma}^\dag\rangle\rangle &=&(1+\frac{u\langle
n^d_{i\bar{\sigma}}\rangle}{\omega+\nu_{\gamma}-u}) (\delta_{ij}-\sum_{b\in
Cu}t_{ib}\langle\langle d_{b\gamma\sigma}\mid
d_{j\lambda\sigma}^\dag\rangle\rangle-\sum_{a\in
Fe,\alpha}t_{ai}'\langle\langle c_{a\alpha\sigma}\mid
d_{j\lambda\sigma}^\dag\rangle\rangle),
\label{Eq:GFM}
\end{eqnarray}
\end{widetext}
with
\begin{eqnarray}
M_\alpha &=&\omega+\mu_\alpha+2\sum_{b\in Cu,\gamma} t_{ib}'(B_\alpha
\langle d_{b\gamma\bar{\sigma}}^\dag c_{i\alpha\bar{\sigma}}\rangle\nonumber\\
&&+\frac{U'-J}{A_\alpha}\sum_{l\neq\alpha,\gamma}\langle
d_{b\gamma\sigma}^\dag c_{il\sigma}\rangle+\frac{U'}{A_\alpha}
\sum_{l\neq\alpha,\gamma}\langle d_{b\gamma\bar{\sigma}}^\dag
c_{il\bar{\sigma}}\rangle),\nonumber\\
N_\alpha &=&1+B_\alpha\langle n_{i\alpha\bar{\sigma}}\rangle
+\frac{U'-J}{A_\alpha}\sum_{l\neq\alpha}\langle
n_{il\sigma}\rangle+\frac{U'}{A_\alpha}
\sum_{l\neq\alpha}\langle n_{il\bar{\sigma}}\rangle,\nonumber\\
A_\alpha &=&\omega +\mu_\alpha-U'\sum_{s\neq\alpha,\sigma}\langle
n_{is\sigma}\rangle+J\sum_{s\neq\alpha}\langle n_{is\sigma}\rangle,\nonumber\\
B_\alpha &=&\frac{U(1+\frac{U'-J}{A_\alpha}
\sum_{l\neq\alpha}\langle n_{il\sigma}\rangle
+\frac{U'}{A_\alpha}\sum_{l\neq\alpha}\langle
n_{il\bar{\sigma}}\rangle)}{\omega +\mu_\alpha
-U-U'\sum_{l\neq\alpha,\sigma}\langle n_{il\sigma}\rangle
+J\sum_{l\neq\alpha}\langle n_{il\sigma}\rangle},\nonumber\\
K_\alpha &=&\frac{U'-J}{A_\alpha}\sum_{l\neq\alpha}\langle
c_{il\sigma}^\dag c_{i\alpha\sigma}\rangle.
\end{eqnarray}

Therefore, the real-space Green's function matrix $\mathscr{G}$
can be obtained self-consistently by the matrix relationship
\begin{equation}
\mathscr{M}\cdot \mathscr{G} = \mathscr{N},
\end{equation}
where $\mathscr{M}$ and $\mathscr{N}$ are also $3N\times 3N$
matrices that consist of the correlation functions, which can be
obtained by the spectral theorem $\langle A B\rangle
=-\frac{1}{\pi}\int_{-\infty}^{+\infty}
f(\omega)\textmd{Im}\langle\langle A\mid B\rangle\rangle $. The
real-space Green's functions can be transformed into the
momentum-space by using the Fourier transformation,
$G_{\textbf{k}}^{\alpha\beta}(\omega)=\frac{1}{N}\sum_{ij}
G_{ij}^{\alpha\beta}(\omega)e^{-i\textbf{k}\cdot(\textbf{R}_i-\textbf{R}_j)}$,
where $\textbf{k}$ denotes the wave-vectors in the unfolded
Brillouin zone (BZ).\cite{Mazin} The one-particle spectral density
is obtained by
\begin{equation}
A(\textbf{k},~\omega)=-\frac{1}{\pi}\textmd{Im}G_{\textbf{k}}(\omega),
\end{equation}
where the matrix of the momentum-space Green's function
$G_{\textbf{k}}(\omega)$ is diagonalized to include the strong
hybridization of $d_{xy}$, $d_{xz}$ and $d_{yz}$ orbitals of Fe
ions. In the next section, we study the effect of disorder
introduced by substituting Fe with Cu in Fe-based superconductors.

To study the substitution effect of Cu ions in a finite square
lattice, a portion of sites chosen randomly should be assigned to the
Cu impurities. For example, if the concentration of the Cu substitution is
$x=0.04$, we choose randomly 16 sites for Cu ions in a $N=20\times 20$
lattice, and all the other sites are for Fe ions. For a certain disorder
configuration, we calculate the real-space Green's function as described
above to obtain physical properties, such as the local density
of states (LDOS) at site $i$ for orbital $\alpha$,
\begin{equation}
\rho_{\alpha}(\textbf{r}_i,\omega)=-\frac{1}{\pi}\textmd{Im}
  G_{ii}^{\alpha\alpha}(\omega).
\end{equation}
To find the averaged DOS of the whole system, we need to calculate
the LDOS of different disorder configurations and then determine the
sample-averaged values. The DOS of Cu and Fe ions can be expressed
respectively as
\begin{eqnarray}
\rho_{\alpha}^{(Fe)}&=&\frac{1}{N_s}\sum_m\frac{1}{N_{Fe}}\sum_{i\in Fe}
\rho_{\alpha}^{(m)}(\textbf{r}_i,\omega)\nonumber\\
\rho_{\alpha}^{(Cu)}&=&\frac{1}{N_s}\sum_m\frac{1}{N_{Cu}}\sum_{i\in Cu}
\rho_{\alpha}^{(m)}(\textbf{r}_i,\omega),
\end{eqnarray}
where $N_{Fe}$ and $N_{Cu}$ represent the total numbers of the Fe and Cu
ions, respectively. $N_s$ is the number of disorder configurations, and
we consider more than one hundred disorder samples in the following
calculations. The effect of disorder
could be studied accurately because our numerical method
is just the same as the real-space exact diagonalization
method\cite{Licciardello} when all the terms of interactions are omitted.


\begin{figure}[htbp]
\includegraphics[scale=0.42]{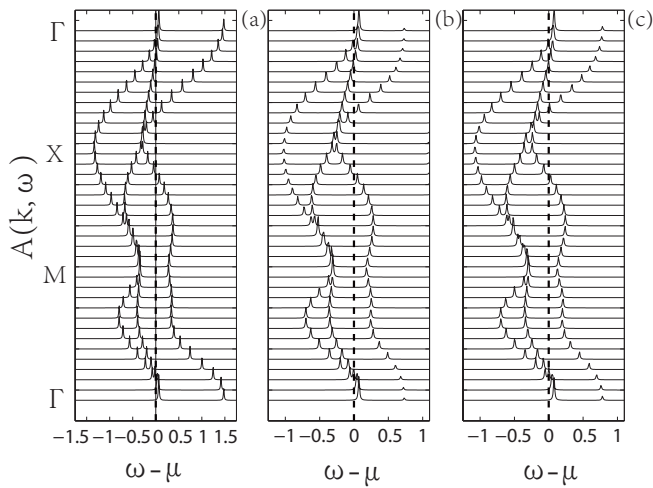}
\caption{Spectral density along the high-symmetry directions in the unfolded
Brillouin zone with different intraorbital interactions: $U=0$ (a),
$U$=1 eV (b), and $U$=1.5 eV (c) for undoped NaFeAs, where the Fermi energy
is indicated by a dashed line. The Hund's rule coupling is $J=U/8$, and $U'=U-2J$
is satisfied.}
\label{fig:SFud}
\end{figure}

\section{Cooperative effect of interactions and disorder}
\label{Sec:U+W}

The strong correlation and disorder, as described above, have been
recognized as two possible fundamental origins that can drive the
MIT. In the correlated fermion systems with the presence of
disorder, the competition between Anderson and Mott-Hubbard MITs is
still a theoretical challenge.\cite{AL50years} Iron-based
superconductors are typically multiorbital correlated materials. The
ARPES measurement reported strong interactions between electrons in
NaFeAs. \cite{HeC} Furthermore, in NaFe$_{1-x}$Cu$_x$As, a strong
impurity potential is introduced by substituting Fe with Cu
randomly. \cite{WangAF} In this section, to elucidate the Cu
substitution effect in iron pnictides doped with copper, we use the
newly developed real-space Green's function method  described above
to study the cooperative effect of the multi-orbital interactions and
doping-induced disorder.

\subsection{Multi-orbital Interactions in undoped NaFeAs}

\begin{figure}[htbp]
\includegraphics[scale=0.35]{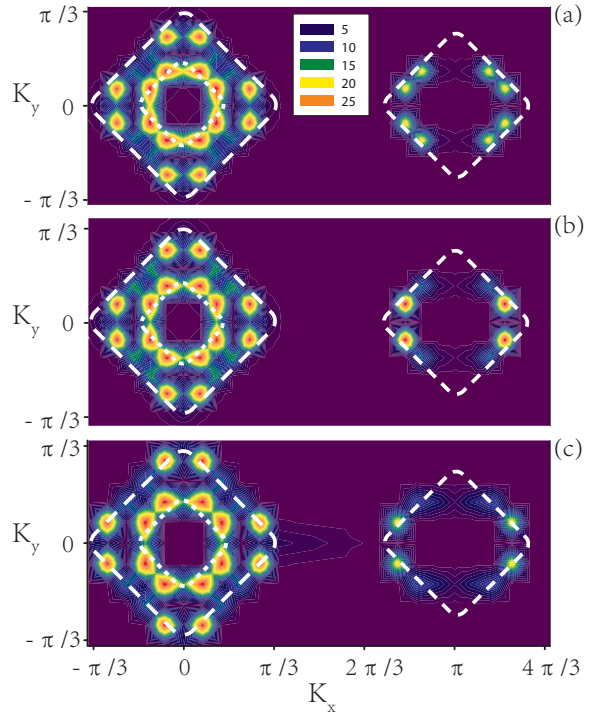}
\caption{(Color online) Fermi surfaces of the undoped three-orbital
Hubbard model with $U=0$ (a), $U$=1 eV (b), and $U$=1.5 eV (c). In an area
of the unfolded Brillouin zone, there are two hole pockets around the $\Gamma$
point (0, 0) and one electron pocket around the X point ($\pi$, 0).
The dashed lines show schematically the Fermi surfaces of the ARPES study on
the parent compound of the 111 phase. \cite{YeZR,Knolle}
The model parameters are the same as those in Fig.~\ref{fig:SFud}.}
\label{fig:EoFS}
\end{figure}

To estimate the multiorbital interaction parameters of Fe ions, such
as the intraorbital interactions $U$, interorbital interactions
$U'$, and Hund's rule coupling $J$, we fit our numerical data of the
band structures to some experimental and theoretical findings, as
shown in Fig.~\ref{fig:SFud}. Because the undoped Fe-based
superconductors are indicated to have a filling of approximately
two-thirds based on the band structure calculations,
\cite{Boeri,Huale} we determine the Fermi energy by using a
constraint of four electrons per Fe,
\begin{equation}
n_{total}=-\frac{1}{\pi}\sum_{i\alpha}\int_{-\infty}^{+\infty}
f(\omega)\textmd{Im} G_{ii}^{\alpha\alpha}(\omega)d\omega\nonumber=4,
\end{equation}
where $f(\omega)$ is the Fermi-Dirac distribution function.

Being consistent with the noninteracting spectral density of
Ref.~[\mycite{Daghofer-2012}], all three $t_{2g}$ bands across the
Fermi surface with bandwidths of $D_1$=1.95 eV, $D_2$=0.9 eV, and
$D_3$=1.5 eV, respectively, are shown in Fig.~\ref{fig:SFud}(a) from
high energy to low energy. A remarkable effect on the spectral
function is found when the multi-orbital interactions are taken into
account, especially when all three bands reduce their bandwidth with
the increasing interactions. According to some previous theoretical
studies of the iron
pnictides,\cite{Daghofer-2012,Daghofer-2010,ZhouS} Hund's rule
coupling is predicted to be $J=U/8$ and the relation $U'=U-2J$ is
also satisfied.
As shown in Fig.~\ref{fig:SFud}(c), the bandwidths of the three
orbitals are $D_1$=1 eV, $D_2$=0.7 eV, and $D_3$=1 eV, respectively,
when $U$ is increased to 1.5 eV. Approximately, the on-site Coulomb
repulsion renormalizes the bandwidth of the three bands across the
Fermi surface by factors of 2, 1.3, and 1.5, respectively, when
$U$=1.5 eV.

Compared with the ARPES measurement,
\cite{YiM,YeZR,Knolle,YangLX-2009,Ding-2008} reasonable
spectral functions and Fermi surfaces are reproduced qualitatively
for undoped Fe-based superconductors NaFeAs and LiFeAs by using the numerical
calculations of the undoped three-orbital Hubbard model. That is,
when $U $= 1.5 eV,  the bandwidths of the spectrum are about the
half of the bandwidths obtained by the LDA, in good agreement with
the ARPES experimental observation.\cite{Ding-2008} Thus, the choice
of the intraorbital interactions as $U$=1.5 eV is reasonable in the
following numerical calculations of the Cu substituting effects in
NaFe$_{1-x}$Cu$_x$As.

Apart from the narrowing of the bandwidths, a significant shift of
the Fermi level is also found accompanying the increasing
interactions. The three-orbital tight-binding model exhibits two
$d_{xz}$/$d_{yz}$-dominant hole pockets around the $\Gamma$ point
and one $d_{xy}$-dominant electron pocket around the X point. The
change of the interactions can only transfer electrons from
$d_{xz}$/$d_{yz}$ orbitals to $d_{xy}$ orbital. With increasing $U$,
a clear evidence of the increasing of the outer hole pocket is shown
in Fig.~\ref{fig:EoFS}. Accordingly, the sizes of the electron
pocket and the inner hole pocket also expand with the increasing
intraorbital interactions, but are not as significant as the size of
the outer hole pocket. Our numerical results are qualitatively
compatible with the LDA+DMFT study on the effects of interactions
in LiFeAs. \cite{LeeG-2012,Ferber-2012}

For the convenience of comparison, we also display schematically
the Fermi surfaces of ARPES study on the parent compound
LiFeAs. \cite{YeZR,Knolle} As shown in Fig.~\ref{fig:EoFS}(a)
and Fig.~\ref{fig:EoFS}(b), both the electron and the outer hole
pockets obtained in the numerical simulations are smaller than the
corresponding experimental results in either case, i.e. when U=0 or
U=1. As shown in Fig.~\ref{fig:EoFS}(c), we can find the best
fitting of the Fermi surfaces obtained numerically with the
experimental results, suggesting that the reasonable value of
the onsite interactions of Fe ions should be $U$=1.5 eV.
Here we need to emphasize that, our theoretical model is applicable
to both the 111 and 122 phases. As indicated by the ARPES
experiments \cite{YeZR}, in LiFeAs and BaFe$_2$As$_2$, the sizes of
the electron and hole pockets are comparable.

\subsection{Effect of Cu substitution on carrier density}

\begin{figure}[htbp]
\includegraphics[scale=0.43]{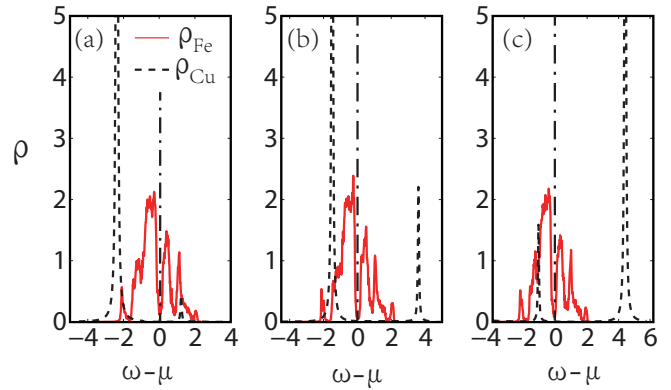}
\caption{(Color online) The average density of states of iron
(solid line) and copper (dashed line)
sites over 100 disorder configurations
for different interactions of Cu ions: $u$=3.6 eV (a),
$u$=5.1 eV (b), and $u$=5.4 eV (c).
The intraorbital interactions of Fe ions are $U=1.5$ eV, and the
concentration of the Cu substitution is $x=0.04$. The lattice size
is $N=20\times 20$, and the total number of disorder samples is
$N_s=100$.}
\label{fig:DOS-FeCu}
\end{figure}

Some experiments concerning the effect of Cu substitution are mainly
carried out in the 111-class and 122-class iron-based
superconductors. In the 111 phase, such as
NaFe$_{1-x}$Cu$_x$As\cite{CuiST} and
LiFe$_{1-x}$Cu$_x$As,\cite{XingLY} it is found that Cu substitution
enlarges the electron pocket and shrinks the hole pockets, implying
that some extra electrons are brought in. Meanwhile, in the
122-phase, the experimental results have not reached agreement yet,
but the prediction of hole doping, introduced by Cu substitution in
122 compounds, is better received. The $3d$ states of a Cu ion in
Ba(Fe$_{1-x}$Cu$_x$)$_2$As$_2$ were found at the bottom of the
valence band, forming a localized 3$d^{10}$ shell.
\cite{YanYJ,WuSF-2015} As a result, the
Cu substitution should result in a hole doping in the 122 phase.
\cite{YanYJ,McLeod,Singh,Anand} Although the volume of the electron
Fermi surface is smaller than the predicted value of the rigid-band
model, effective electron doping introduced by the Cu substitutions
was also predicted by other experiments.\cite{Ideta,ChengP}

\begin{figure}[htbp]
\includegraphics[scale=0.43]{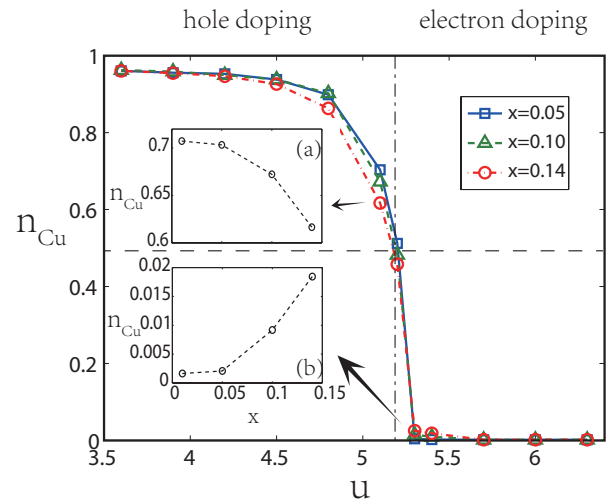}
\caption{(Color online) The average carrier occupancy
on each copper site as a function of the interactions $u$.
Inset: The substitution dependence of $n_{Cu}$ for different
interactions: $u$=5.1 eV (a), and $u$=5.4 eV (b) when
$U$=1.5 eV and $J=U/8$. The lattice
sizes are $N=20\times 20$, and one hundred disorder
configurations ($N_s=100$) are considered.}
\label{fig:nCu}
\end{figure}

Next, we discuss the distinct effects of Cu substitutions on the
carrier density in NaFe$_{1-x}$Fe$_x$As and
Ba(Fe$_{1-x}$Cu$_x$)$_2$As$_2$. A first-principles calculation has
indicated that the screened Coulomb interaction at Fe atoms of 122
families is less than that of Fe atoms of 111 families for
approximately 0.35 eV. \cite{Miyake}  As a result, for the two
types of iron-based superconductors, the onsite interactions of the
Cu impurities should be different accordingly.
Because of the comparative localization of the Cu orbitals,
the effective interactions of Cu ions should be considerably large.
By contrast, the wave functions of Fe orbitals that cross the Fermi
level are extended, introducing small effective interactions of Fe ions.
For simplicity, in our calculation we assume that
the interactions at Cu sites ($u$) are in direct proportion to the
interactions at Fe atoms ($U$), with an approximate relation $u\simeq 4U$.
Considering that in Fe-based superconductors, the value of $U$ should be
1 eV$<U<$2 eV, it is reasonable to predict the interactions of Cu ion
as $u\approx$ 4-7 eV. The above estimated parameters have been proved
to be appropriate by
a combined valence band photoemission and Auger spectroscopy study of
single crystalline Ba(Fe$_{1-x}$Cu$_x$)$_2$As$_2$, where the interactions
in Fe and Cu ions are found to be (1.4$\pm$0.6) eV and (7.5$\pm$0.4) eV,
respectively. \cite{Kraus} Assuming that $u\simeq 4U$, the deviation is approximately
1.4 eV for the Cu interactions of the 111 and 122 phases.

Effects of the onsite interactions of Cu ions on the DOS are shown in
Fig.~\ref{fig:DOS-FeCu}, where the concentration of Cu substitutions is
fixed at $x=0.04$. For simplicity, the interactions of Fe ions are also
fixed as $U$=1.5 eV and $J=U/8$.
When $u=3.6$ eV, we find very narrow upper and lower
Hubbard bands of Cu impurity that locate at $\omega=0.7$ eV
and $\omega=-2.4$ eV respectively, as shown in
Fig.~\ref{fig:DOS-FeCu}(a). Furthermore, the DOS of lower Hubbard band is
considerably larger than that of the upper Hubbard band, suggesting
that most of the electrons in Cu ions locate far below the Fe bands
when $u=3.6$ eV. We also find that the DOS of the copper sites shifts
from the lower Hubbard band to the upper band with increasing $u$ as
shown in Fig.~\ref{fig:DOS-FeCu}(b) and (c). In addition, the upper
Hubbard band moves upwards, such as the upper Hubbard band raises to
$\omega$=4.3 eV when $u$=5.4 eV, which is far above the Fe
bands. Our numerical results demonstrate that electrons in the
Cu band can transfer to the Fe bands with the enhancement of the
interactions of Cu ions.

As shown in Fig.~\ref{fig:nCu}, the carrier occupancy on Cu sites is
obtained by integrating the corresponding DOS of Cu ions. We find
that the average electron occupancy on Cu sites decreases with
increasing $u$, especially the curve of $n_{\textmd{\tiny{Cu}}}$ vs
$u$, which drops sharply in the vicinity of $u$=5.2 eV. From the
perspective of the carrier density on the Fe sites, the Cu
substitutions lead to a hole doping when $u<$5.2 eV because most
electrons rest at the Cu orbitals. However, it is entirely different
when $u\geq$5.2 eV, where the carrier occupancy of the Cu sites
drops to zero, leading to electron doping.

To make it more clear,
the inset figures in Fig.~\ref{fig:nCu} show a reversal tendency of
the dependence of $n_{\textmd{\tiny{Cu}}}$ on $x$ for different $u$.
It is shown that the electron occupancy of each Cu site decreases
with increasing $x$ within the hole doping region with $u$=5.1 eV.
On the contrary, the occupancy increases with increasing $x$ in the
electron doping region with $u$=5.4 eV. The ARPES experiment on
NaFeAs\cite{CuiST} indicates that the total number of electron
carriers introduced by Cu atoms increases when the Cu substitution
increases from 0.019 to 0.14. However, the average number of
electrons in Cu atom decreases significantly from 1.9 to 0.64 with
increasing doping concentration. From the inset in
Fig.~\ref{fig:nCu}(b), we find that the average electron occupancy
on a single Cu site also increases with increasing $x$, implying
that less electrons can be pushed from Cu sites to the iron sites.
This tendency is consistent with the ARPES experiment.

\begin{figure}[htbp]
\includegraphics[scale=0.5]{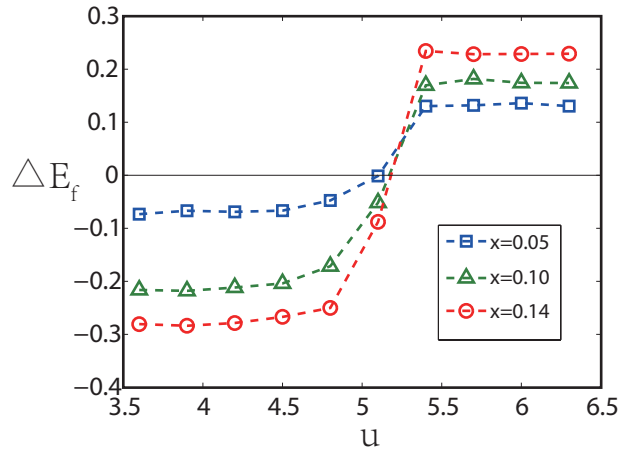}
\caption{(Color online) The variation of E$_f$, $\Delta$E$_f=E_f-E_{f0}$,
vs the interactions $u$ of Cu ions,
where $E_{f0}$ is the Fermi energy in undoped compound. The lattice
parameters are the same as those in Fig.~\ref{fig:nCu}.}
\label{fig:Ef-Cu}
\end{figure}

In Fig.~\ref{fig:Ef-Cu}, we plot the dependence of the variation of the
Fermi energy E$_f$ on the interactions $u$. In the weak interaction
range of $u$, the Fermi energy of a Cu-substituted compound is less
than that of an undoped compound, being consistent with the hole
doping feature. However, when $u>$5.2 eV, $\Delta$E$_f$ becomes
positive, which means an electron doping appears. Our calculations
demonstrate that the Coulomb interactions $u$ affect the average
carrier density on iron sites and the Fermi energy of the whole
system by pushing the electrons out of copper orbitals. For this
reason, the different effects of Cu substitutions on Fermi energy
and carrier density in 111 and 122 phases can be understood because
the correlations of the 122 families are smaller than the
correlations of the 111 families.

\begin{figure}[htbp]
\includegraphics[scale=0.38]{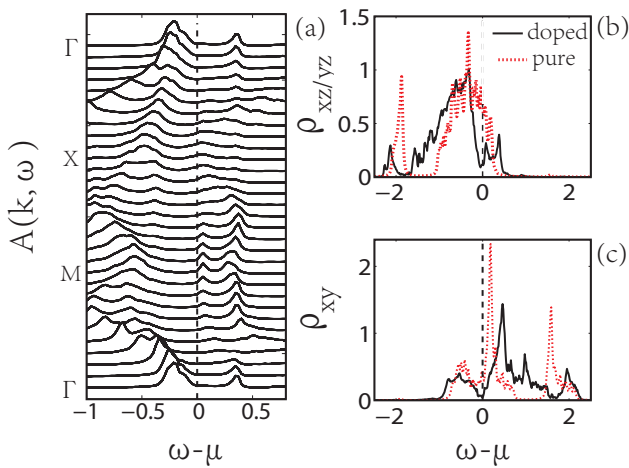}
\caption{(Color online) (a) Spectral density of Cu-doped NaFeAs with
Cu impurity concentration $x=0.04$. (b) and (c) show the density of
states of the degenerate $d_{xz}$/$d_{yz}$ and $d_{xy}$ orbitals for
$x=0$ and $x=0.04$, respectively. The Fermi energy level is
represented by the black dashed line. The lattice size is $N=20\times 20$,
and the number of disorder configurations is $N_s=100$.
The other parameters are $U$=1.5 eV,
$J=U/8$, and $u$=6 eV.}
\label{fig:ESgap-Cu}
\end{figure}

\subsection{Localization effects of Cu substitutions}

Aside from the change in carrier density, another remarkable effect
of the Cu substitutions is the doping-induced disorder. Adding the
Cu substitutions, the bands are most strongly modified near the
Fermi level. It is shown that both two hole pockets around the
$\Gamma$ point and one electron pocket around the $X$ point are very
sensitive to the Cu substitution. Apart from the shrinkage of the
bandwidth, the most dramatic change in the band structures in
NaFe$_{1-x}$Cu$_x$As is that the spectral density near the Fermi
surface is strongly suppressed, as shown in
Fig.~\ref{fig:ESgap-Cu}(a). As a result, the compound exhibits
insulating or bad metallic behavior.

Efros and Shklovskii have demonstrated that the interactions between
the localized electrons in a disordered system can create a Coulomb
gap in the DOS near the Fermi level.\cite{Efros} As a result, the
appearance of the zero-bias anomaly at the Fermi energy is seen as
proof of the localization of electronic states in the correlated
systems.\cite{Song} Just as expected, a soft gap can be found at the
Fermi level in the DOS of both the degenerated $d_{xz}/d_{yz}$
orbitals and the $d_{xy}$ orbital when $x=0.04$, as shown in
Fig.~\ref{fig:ESgap-Cu}(b) and (c). The characteristic of the DOS of
undoped compound is entirely different because there is no soft gap
at Fermi level for all three orbitals, as shown by the red dashed lines in
Fig.~\ref{fig:ESgap-Cu}(b) and (c). Moreover, the Efros-Shklovskii
gap can also be found in Fig.~\ref{fig:DOS-FeCu}. Therefore, we believe
that the disorder effect in NaFe$_{1-x}$Cu$_x$As is considerably
strong.

\begin{figure}[htbp]
\includegraphics[scale=0.33]{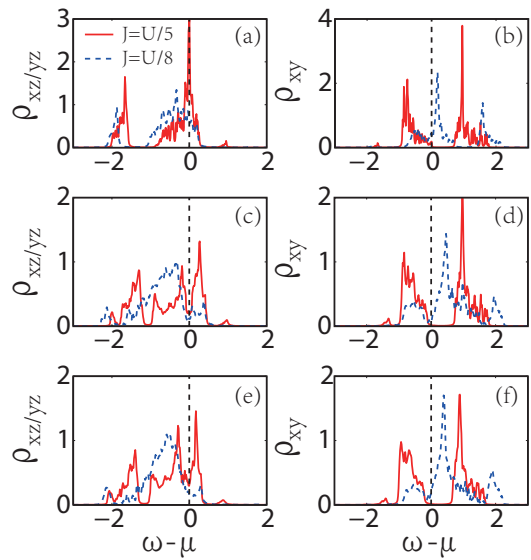}
\caption{(Color online) The evolution of the density of state of
the $d_{xy}$ orbital (the left panel) and the $d_{xz}/d_{yz}$
orbitals (the right panel) for different Cu impurity
concentrations: $x=0$ ((a) and (b)), $x=0.02$ ((c) and (d)),
and $x=0.08$ ((e) and (f)) when $J=U/8$ (dashed line) and
$J=U/5$ (solid line), respectively. The other parameters
are: $U$=1.5 eV and $u$=6 eV.
The lattice sizes are $N=20\times 20$, and the number of disorder
samples is $N_s=100$.}
\label{fig:AM-Cu}
\end{figure}

The iron-based superconductors are predicted to be close to the MIT
border but are in the metallic side.\cite{Huale} There are three
possible origins for the MIT in the correlated system with disorder:
(1) Mott-Hubbard MIT: the on-site Coulomb interactions split the
energy band near the Fermi level into higher and lower Hubbard
bands; (2) band occupation-driven MIT: changing the carrier density
or bandwidth to make the conducting band fully filled; (3) Anderson
MIT: electrons are localized in the real-space by Anderson
localization effects introduced by disorder. Experiments found Cu
substitution-driven MIT in NaFe$_{1-x}$Cu$_x$As, but the mechanism
of the MIT is still on debatable.

When $U$=1.5 eV and $J$=U/8, it is found that Anderson MIT happens
with the increasing of the Cu substitutions, owing to the effect of
disorder introduced by the Cu dopant in Fe-based superconductors.
Besides, there is no Mott gap for both the $d_{xy}$ and
$d_{xz}/d_{yz}$ orbitals at Fermi level when $J=U/8$,
as shown in Fig.~\ref{fig:AM-Cu}.
In general, the multi-orbital interactions, especially the Hund's
rule couplings are likely to cause orbital-selective
transitions.\cite{Medici-2009} It has also been predicted
in some theoretical studies that a wide range of $J/U$
is possible for the iron-based superconductors. For example,
the proper range of $J/U$ couplings was predicted as 0.1$<J/U<$0.33,
which was obtained by comparing the theoretical results
with neutron scattering and ARPES experiments.\cite{LuoQL}
Therefore, we also study the effect of the Hund's rule coupling
on the MIT in the Cu-substituted iron pnictides.
An orbital-selective insulating phase is
found when we increase the Hund's rule coupling
to $J=U/5$ and the on-site interactions of Cu ions to $u=4U$=6 eV.
By observing the evolution of the DOS of the three $3d$ orbitals,
we find that the properties of the insulating phase of the
$d_{xz}/d_{yz}$ and $d_{xy}$ orbitals are different. As shown in the
left panel of Fig.~\ref{fig:AM-Cu}, a zero-bias anomaly dip can be
found with increasing Cu substitution in the DOS of the
$d_{xz}/d_{yz}$ orbital, but no hard gap can be obtained. Therefore,
the lower-energy states at the Fermi level are Anderson localized as
a result. Conversely, as shown in the right panel of
Fig.~\ref{fig:AM-Cu}, a hard Mott gap is opened in the Fermi level,
indicating that the $d_{xy}$ orbital is in the Mott insulating
state. In addition, when $J=U/5$, the carrier type also changes
from a hole doping to an electron doping with
the increasing of the Coulomb interaction $u$ of Cu ions.

The effect of disorder on the conductance of Fe-based
superconductors with Cu doping is studied by using the Kubo formula,
\cite{Kubo,PALee} which can be simplified, when $\omega=0$, as a
$3\times 3$ matrix $\sigma$ for the three orbital Hubbard model,
\begin{eqnarray}
\sigma&=&\frac{e^2}{\pi} \textmd{Tr}
[\textmd{Im}G_i(j,j')\textmd{Im}G_i(j'-1,j-1)\nonumber\\
&&+\textmd{Im}G_i(j-1,j'-1)\textmd{Im}G_i(j',j)\nonumber\\
&&-\textmd{Im}G_i(j,j'-1)\textmd{Im}G_i(j',j-1)\nonumber\\
&&-\textmd{Im}G_i(j-1,j')\textmd{Im}G_i(j'-1,j)].
\label{Eq:Cond}
\end{eqnarray}
The trace is over the sites perpendicular to the current direction
($k$ direction). ($i$, $j$) denote the coordinates of lattice sites
in the ribbon, where the periodic boundary conditions should be
applied only to the $i$ direction, and $j$ and $j'$ are chosen to be
on opposite sides of the sample along the current direction. The
conductance of the whole system can be obtained by adding the
conductance of all three orbitals, which correspond to the three
diagonal elements of the conductance matrix $\sigma$. As mentioned
in Sec.~\ref{Sec:RSGF}, the averaged conductance is extracted by
\begin{equation}
\bar{\sigma}=\frac{1}{N_s}\sum_m\sigma^{(m)},
\end{equation}
where $\sigma^{(m)}$ is the conductance obtained by Eq.~(\ref{Eq:Cond})
for the m-$th$ disorder configuration, and the resistance could be
obtained directly by $\rho=1/\bar{\sigma}$.

\begin{figure}[htbp]
\includegraphics[scale=0.43]{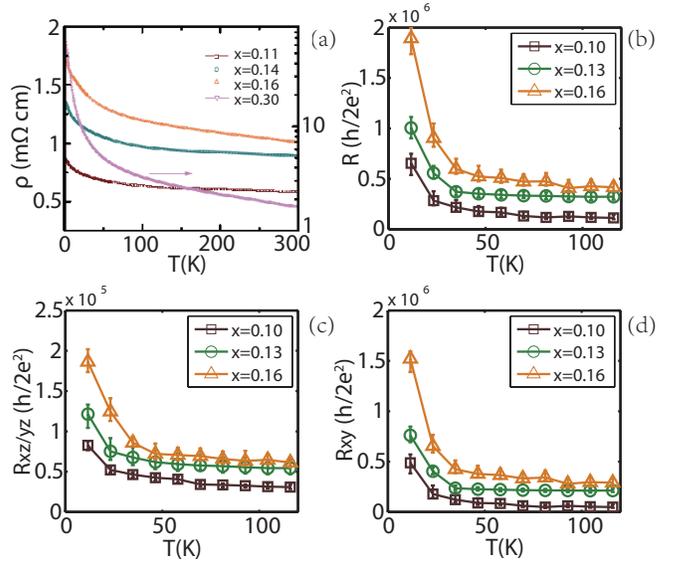}
\caption{(Color online) The temperature dependence of the resistance with
increasing Cu substitution from $x=0.1$ to $x=0.16$.
(a) Experimental data taken from Ref. [60]. (b) Numerical result
of the resistance for the whole system. Additionally, the resistance of
each orbital  for $d_{xz}$/$d_{yz}$ (c) and $d_{xy}$ (d). Each data
point is obtained by averaging for one hundred disorder configurations.
The lattice sizes are $N=20\times 20$, and one hundred disorder
configurations ($N_s=100$) are considered.
The other model parameters are the same as in Fig.~\ref{fig:AM-Cu}.
}
\label{fig:Cond}
\end{figure}

To compare with the experimental results, in Fig.~\ref{fig:Cond}, we
show the temperature dependence of the resistance for different Cu
concentrations.  It is obvious that the compound behaves as an
insulator when the density of the Cu substitution is over $x=0.1$.
We find that the MIT in NaFe$_{1-x}$Cu$_x$As is a transition from a
metal to an orbital-selective insulating phase, which is caused by
the cooperative effect of multi-orbital interactions and the
doping-induced disorder. With the increasing of Cu substitutions,
the effect of disorder significantly enhances, leading to strong
Anderson localization in the $d_{xz}$ and $d_{yz}$ orbitals.
Meanwhile, a Mott gap occurs in the $d_{xy}$ orbital. As a result,
the resistance increases with increasing doping concentration both
for the whole system and the individual orbitals, in good agreement
with the experimental results. \cite{Thesis}

\section{Conclusion}
\label{Sec:Con}

We present the study of an extended real-space Green's function
method combined with density function calculation on the
cooperative effects of multiorbital correlations and
disorder in Cu-substituted iron-based superconductors.
The inhomogeneous three-orbital Hubbard
model has been employed to describe the substitution of Cu with Fe
based on the electron structures of NaFe$_{1-x}$Cu$_x$As compounds.
We find that the type of charge carriers introduced by the Cu
substitution, namely, electrons or holes, is virtually determined
by the effective interactions of the copper atoms. Due to the
different environments surrounding the Cu substitution, the interactions
of Cu ions in 111 phase are larger than that of 122 phase. As a result,
a hole doping is introduced by the Cu substitution in
Ba(Fe$_{1-x}$Cu$_{x}$)$_2$As$_2$, as opposed to an electron doping in
NaFe$_{1-x}$Cu$_x$As. Moreover, an orbital-selective
insulating phase is observed where the $d_{xz}$ and $d_{yz}$
orbitals are Anderson localized, while the $d_{xy}$ orbital is in
the Mott insulating phase. These results explain the
experimental observations in NaFe$_{1-x}$Cu$_x$As and
Ba(Fe$_{1-x}$Cu$_{x}$)$_2$As$_2$.

\section*{Acknowledgments}
The computational resources utilized in this research were provided
by Shanghai Supercomputer Center. The work was supported by the NSFC
of China, under Grants No. 10974018, No. 11174036, No. 11474023, and
No. 11274310, the National Basic Research Program of China (Grant No.
2011CBA00108), and the Fundamental Research Funds for the Central
Universities.


\begin{table*}[htb]
\caption{\label{Tb_Hop} The NN and NNN hopping integrals of the
electrons in the three orbitals $d_{xy}$, $d_{xz}$ and $d_{yz}$ of
Fe ions. $(-1)^{|i|}$ indicates that the parameters change sign
along the site locations.} \
\begin{tabular}{c|rrrrrrrrr}
\hline \hline \\ [-4pt]
  T$_{i\alpha,j\beta}$  \\ [+2pt]
\hline \\ [-4pt]
 $(\alpha, \beta)$ $\backslash$ $|\bm{R}_i-\bm{R}_j|$
 & \big[$1,0$\big]
 & \big[$0,1$\big]
 & \big[$-1,0$\big]
 & \big[$0,-1$\big]
 & \big[$1,1$\big]
 & \big[$-1,1$\big]
 & \big[$-1,-1$\big]
 & \big[$1,-1$\big]
\\ [+4pt]
 \hline \\ [-8pt]             
 $(xy,xy)$                    &              $T_1$ &              $T_1$ &                $T_1$ &                $T_1$ &                $T_2$ &                $T_2$ &              $T_2$ &                $T_2$ &  \\ [+4pt]
 $(xy,xz)$                    &  ~~$(-1)^{|i|}T_3$ &                  0 &  ~~$(-1)^{|i+1|}T_3$ &                    0 &  ~~$(-1)^{|i+1|}T_4$ &    ~~$(-1)^{|i|}T_4$ &  ~~$(-1)^{|i|}T_4$ &  ~~$(-1)^{|i+1|}T_4$ &  \\ [+4pt]
 $(xy,yz)$                    &                  0 &  ~~$(-1)^{|i|}T_3$ &                    0 &  ~~$(-1)^{|i+1|}T_3$ &  ~~$(-1)^{|i+1|}T_4$ &  ~~$(-1)^{|i+1|}T_4$ &  ~~$(-1)^{|i|}T_4$ &    ~~$(-1)^{|i|}T_4$ &  \\ [+4pt]
 $(xz,xz)$                    &              $T_6$ &              $T_5$ &                $T_6$ &                $T_5$ &                $T_7$ &                $T_7$ &              $T_7$ &                $T_7$ &  \\ 
 $(yz,yz)$                    &              $T_5$ &              $T_6$ &                $T_5$ &                $T_6$ &                $T_7$ &                $T_7$ &              $T_7$ &                $T_7$ &  \\ [+4pt]
 $(yz,xz)$                    &                  0 &                  0 &                    0 &                    0 &               $-T_8$ &                $T_8$ &             $-T_8$ &                $T_8$ &  \\ [+4pt]
\hline \hline
\end{tabular}
\end{table*}

\appendix
\section{Hopping integrals in the three-orbital tight-binding model}
\label{APP-HT}

As shown in Figs.~\ref{fig:BS-FeCu}-~\ref{fig:DOSLDA-Cu}, the local density function
calculation predicted that the Fermi surface of NaFe$_{1-x}$Cu$_x$As
composed only three Fe $3d$ orbitals ($d_{xy}$, $d_{xz}$ and
$d_{yz}$), whereas Cu $3d$ orbitals contributed from -4 eV to -2 eV
under the Fermi level. (Eq.~\ref{Eq_HamFe}). Daghofer $et$ $al.$
have constructed a three-orbital Hamiltonian for the pnictides
following the Slater-Koster
procedure.\cite{Daghofer-2010,Daghofer-2012} In keeping with their
model, we mainly consider the NN and
next-nearest-neighbor (NNN) hopping integrals of electrons in the
three orbitals $d_{xy}$, $d_{xz}$, and $d_{yz}$ of iron ions. The
definitions of the different types of hopping terms
$T_{i\alpha,j\beta}$ in the tight-binding model are displayed in
Table~\ref{Tb_Hop}.

The hopping integrals between the $d_{xy}$ and $d_{xz}$/$d_{yz}$
orbitals contain factors $(-1)^{i}$ that arise from the two-iron
unit cell of the original FeAs planes.\cite{Daghofer-2010} As a
result, those interorbital hopping terms change sign for NN iron
sites. Moreover, the NN intraorbital hoppings between $d_{xy}$
and $d_{yz}$ along the $\hat{x}$ direction, and between $d_{xy}$
and $d_{xz}$ along the $\hat{y}$ direction, as well as the NN
interorbital hopping between $d_{xz}$ and $d_{yz}$ orbitals, are all
omitted.

Apart from the hopping terms, we need to add a term in the
Hamiltonian for the energy splitting between the $d_{xy}$ and
degenerate $d_{xz}$/$d_{yz}$ orbitals, $\Delta_{xy}$, which is
crucial to providing the proper Fermi pockets when fitting with the
local density approximation results.

\section{Decoupling approach}
\label{APP-DA}

All approximations that we used to decouple the third-order Green's
functions to obtain Eq.~(\ref{Eq:GFM}) are listed as follows:
\begin{eqnarray}
\langle\langle n_{i\alpha\sigma'}c_{bm\sigma}\mid
c_{j\beta\sigma}^\dag\rangle\rangle&\approx& \langle
n_{i\alpha\sigma'}\rangle\langle\langle c_{bm\sigma}\mid
c_{j\beta\sigma}^\dag\rangle\rangle,
  \nonumber\\
\langle\langle n_{i\alpha\sigma'}d_{b\gamma\sigma}\mid
c_{j\beta\sigma}^\dag\rangle\rangle&\approx& \langle
n_{i\alpha\sigma'}\rangle\langle\langle d_{b\gamma\sigma}\mid
c_{j\beta\sigma}^\dag\rangle\rangle,
  \nonumber\\
\langle\langle c_{bm\bar{\sigma}}^\dag
c_{i\alpha\bar{\sigma}}c_{i\alpha\sigma}\mid
c_{j\beta\sigma}^\dag\rangle\rangle&\approx& \langle
c_{bm\bar{\sigma}}^\dag c_{i\alpha\bar{\sigma}}\rangle\langle\langle
c_{i\alpha\sigma}\mid
c_{j\beta\sigma}^\dag\rangle\rangle,
  \nonumber\\
\langle\langle c_{i\alpha\bar{\sigma}}^\dag
c_{bm\bar{\sigma}}c_{i\alpha\sigma}\mid
c_{j\beta\sigma}^\dag\rangle\rangle&\approx& \langle
c_{i\alpha\bar{\sigma}}^\dag c_{bm\bar{\sigma}}\rangle\langle\langle
c_{i\alpha\sigma}\mid
c_{j\beta\sigma}^\dag\rangle\rangle,
  \nonumber\\
\langle\langle c_{bm\bar{\sigma}}^\dag
c_{il\bar{\sigma}}c_{i\alpha\sigma}\mid
c_{j\beta\sigma}^\dag\rangle\rangle&\approx& \langle
c_{bm\bar{\sigma}}^\dag c_{il\bar{\sigma}}\rangle\langle\langle
c_{i\alpha\sigma}\mid
c_{j\beta\sigma}^\dag\rangle\rangle,
  \nonumber\\
\langle\langle c_{il\bar{\sigma}}^\dag
c_{bm\bar{\sigma}}c_{i\alpha\sigma}\mid
c_{j\beta\sigma}^\dag\rangle\rangle&\approx& \langle
c_{il\bar{\sigma}}^\dag c_{bm\bar{\sigma}}\rangle\langle\langle
c_{i\alpha\sigma}\mid
c_{j\beta\sigma}^\dag\rangle\rangle,
  \nonumber\\
\langle\langle c_{bm\sigma}^\dag c_{il\sigma}c_{i\alpha\sigma}\mid
c_{j\beta\sigma}^\dag\rangle\rangle&\approx& \langle
c_{bm\sigma}^\dag c_{il\sigma}\rangle\langle\langle
c_{i\alpha\sigma}\mid c_{j\beta\sigma}^\dag\rangle\rangle\nonumber\\
&&-\langle c_{bm\sigma}^\dag c_{i\alpha\sigma}\rangle\langle\langle
c_{il\sigma}\mid c_{j\beta\sigma}^\dag\rangle\rangle,
 \nonumber\\
\langle\langle c_{il\sigma}^\dag c_{bm\sigma}c_{i\alpha\sigma}\mid
c_{j\beta\sigma}^\dag\rangle\rangle&\approx& \langle
c_{il\sigma}^\dag c_{bm\sigma}\rangle\langle\langle
c_{i\alpha\sigma}\mid
c_{j\beta\sigma}^\dag\rangle\rangle\nonumber\\
&&-\langle c_{il\sigma}^\dag c_{i\alpha\sigma}\rangle\langle\langle
c_{bm\sigma}\mid c_{j\beta\sigma}^\dag\rangle\rangle,
 \nonumber\\
\langle\langle d_{b\gamma\bar{\sigma}}^\dag
c_{i\alpha\bar{\sigma}}c_{i\alpha\sigma}\mid
c_{j\beta\sigma}^\dag\rangle\rangle&\approx& \langle
d_{b\gamma\bar{\sigma}}^\dag c_{i\alpha\bar{\sigma}}\rangle\langle\langle
c_{i\alpha\sigma}\mid
c_{j\beta\sigma}^\dag\rangle\rangle,
 \nonumber\\
\langle\langle c_{i\alpha\bar{\sigma}}^\dag
d_{b\gamma\bar{\sigma}}c_{i\alpha\sigma}\mid
c_{j\beta\sigma}^\dag\rangle\rangle&\approx& \langle
c_{i\alpha\bar{\sigma}}^\dag d_{b\gamma\bar{\sigma}}\rangle\langle\langle
c_{i\alpha\sigma}\mid
c_{j\beta\sigma}^\dag\rangle\rangle,
 \nonumber\\
\langle\langle d_{b\gamma\bar{\sigma}}^\dag
c_{il\bar{\sigma}}c_{i\alpha\sigma}\mid
c_{j\beta\sigma}^\dag\rangle\rangle&\approx& \langle
d_{b\gamma\bar{\sigma}}^\dag c_{il\bar{\sigma}}\rangle\langle\langle
c_{i\alpha\sigma}\mid
c_{j\beta\sigma}^\dag\rangle\rangle,
 \nonumber\\
\langle\langle c_{il\bar{\sigma}}^\dag
d_{b\gamma\bar{\sigma}}c_{i\alpha\sigma}\mid
c_{j\beta\sigma}^\dag\rangle\rangle&\approx& \langle
c_{il\bar{\sigma}}^\dag d_{b\gamma\bar{\sigma}}\rangle\langle\langle
c_{i\alpha\sigma}\mid
c_{j\beta\sigma}^\dag\rangle\rangle,
 \nonumber\\
\langle\langle d_{b\gamma\sigma}^\dag c_{il\sigma}c_{i\alpha\sigma}\mid
c_{j\beta\sigma}^\dag\rangle\rangle&\approx& \langle
d_{b\gamma\sigma}^\dag c_{il\sigma}\rangle\langle\langle
c_{i\alpha\sigma}\mid
c_{j\beta\sigma}^\dag\rangle\rangle\nonumber\\
&&-\langle d_{b\gamma\sigma}^\dag c_{i\alpha\sigma}\rangle\langle\langle
c_{il\sigma}\mid c_{j\beta\sigma}^\dag\rangle\rangle,
  \nonumber\\
\langle\langle c_{il\sigma}^\dag d_{b\gamma\sigma}c_{i\alpha\sigma}\mid
c_{j\beta\sigma}^\dag\rangle\rangle&\approx& \langle
c_{il\sigma}^\dag d_{b\gamma\sigma}\rangle\langle\langle
c_{i\alpha\sigma}\mid
c_{j\beta\sigma}^\dag\rangle\rangle\nonumber\\
&&-\langle c_{il\sigma}^\dag c_{i\alpha\sigma}\rangle\langle\langle
d_{b\gamma\sigma}\mid c_{j\beta\sigma}^\dag\rangle\rangle,
  \nonumber\\
\langle\langle
n_{il\sigma'}n_{i\alpha\bar{\sigma}}c_{i\alpha\sigma}\mid
c_{j\beta\sigma}^\dag\rangle\rangle&\approx& \langle
n_{il\sigma'}\rangle\langle\langle
n_{i\alpha\bar{\sigma}}c_{i\alpha\sigma}\mid
c_{j\beta\sigma}^\dag\rangle\rangle,
 \nonumber\\
\langle\langle n_{is\sigma'}n_{il\sigma''}c_{i\alpha\sigma}\mid
c_{j\beta\sigma}^\dag\rangle\rangle&\approx& \langle
n_{is\sigma'}\rangle\langle\langle
n_{il\sigma''}c_{i\alpha\sigma}\mid
c_{j\beta\sigma}^\dag\rangle\rangle\nonumber\\
&&+\langle n_{il\sigma''} \rangle\langle\langle
n_{is\sigma'}c_{i\alpha\sigma}\mid
c_{j\beta\sigma}^\dag\rangle\rangle,
 \nonumber
\end{eqnarray}
where $\sigma'$ and $\sigma''$ represent $\sigma$ or $\bar{\sigma}$.

In the above decoupling scheme, we follow the standard decoupling
method \cite{Nolting} and introduce three basic rules
for the multi-orbital model as: (1) Some specific second-order
Green's functions, such as $\langle\langle
n_{i\alpha\bar{\sigma}}c_{i\alpha\sigma}\mid
c_{j\beta\sigma}^\dag\rangle\rangle$, $\langle\langle
n_{il\sigma}c_{i\alpha\sigma}\mid
c_{j\beta\sigma}^\dag\rangle\rangle$, and $\langle\langle
n_{il\bar{\sigma}}c_{i\alpha\sigma}\mid
c_{j\beta\sigma}^\dag\rangle\rangle$, are regarded as basic Green's
functions that receive the same treatment as the first-order Green's
function. That is, we also need to find the equations of motions of
those Green's functions. (2) In a third-order Green's function with
$n$ operators, such as $\langle\langle
n_{il\sigma'}n_{i\alpha\bar{\sigma}} c_{i\alpha\sigma}\mid
c_{j\beta\sigma}^\dag\rangle\rangle$, we introduce the average value
of $\langle n_{il\sigma'}\rangle$ to get the lower-order Green's
function as $\langle n_{il\sigma'}\rangle\langle\langle
n_{i\alpha\bar{\sigma}}c_{i\alpha\sigma}\mid
c_{j\beta\sigma}^\dag\rangle\rangle$ because its orbital index $l$
is different from $\alpha$ of the other operators. (3) We do not
consider the effects of the correlation functions that contain spin
flip terms. As a result, the equations of motion of some basic
second-order Green's functions can also be obtained as
\begin{widetext}
\begin{eqnarray}
(\omega +\mu_\alpha &-&U_1-U'\sum_{l\neq\alpha}\langle
n_{il}\rangle+J_z\sum_{l\neq\alpha}\langle
n_{il\sigma}\rangle)\langle\langle
n_{i\alpha\bar{\sigma}}c_{i\alpha\sigma}\mid
c_{j\beta\sigma}^\dag\rangle\rangle =\langle
n_{i\alpha\bar{\sigma}}\rangle\delta_{ij}\delta_{\alpha\beta}
-\sum_{b\in Fe,m}T_{ib\alpha m}\langle
n_{i\alpha\bar{\sigma}}\rangle\langle\langle c_{bm\sigma}\mid
c_{j\beta\sigma}^\dag\rangle\rangle
 \nonumber\\
&&-\sum_{b\in Cu}t_{ib}'(2\langle d_{b\bar{\sigma}}^\dag
c_{i\alpha\bar{\sigma}}\rangle\langle\langle c_{i\alpha\sigma}\mid
c_{j\beta\sigma}^\dag\rangle\rangle
+\langle n_{i\alpha\bar{\sigma}}\rangle\langle\langle
d_{b\sigma}\mid c_{j\beta\sigma}^\dag\rangle\rangle),\\
(\omega +\mu_\alpha &-&U'\sum_{s\neq\alpha}\langle
n_{is}\rangle+J_z\sum_{s\neq\alpha}\langle
n_{is\sigma}\rangle)\langle\langle n_{il\sigma}c_{i\alpha\sigma}\mid
c_{j\beta\sigma}^\dag\rangle\rangle
=\langle n_{il\sigma}\rangle\delta_{ij}\delta_{\alpha\beta}-
\langle c_{il\sigma}^\dag c_{i\alpha\sigma}\rangle\delta_{ij}\delta_{l\beta}
 \nonumber\\
&&-\sum_{b\in Fe,m}T_{iblm}(\langle c_{bm\sigma}^\dag
c_{i\alpha\sigma}\rangle\langle\langle c_{il\sigma}\mid
c_{j\beta\sigma}^\dag\rangle\rangle -\langle c_{il\sigma}^\dag
c_{i\alpha\sigma}\rangle\langle\langle c_{bm\sigma}\mid
c_{j\beta\sigma}^\dag\rangle\rangle)
-\sum_{b\in Fe,m}T_{ib\alpha m}\langle n_{il\sigma}\rangle\langle\langle
c_{bm\sigma}\mid c_{j\beta\sigma}^\dag\rangle\rangle
 \nonumber\\
&&-\sum_{b\in Cu}t_{ib}'[(\langle n_{il\sigma}\rangle-\langle
c_{il\sigma}^\dag c_{i\alpha\sigma}\rangle)\langle\langle
d_{b\sigma}\mid c_{j\beta\sigma}^\dag\rangle\rangle) +2\langle
d_{b\sigma}^\dag c_{il\sigma}\rangle\langle\langle
c_{i\alpha\sigma}\mid c_{j\beta\sigma}^\dag\rangle\rangle
-\langle d_{b\sigma}^\dag c_{i\alpha\sigma}\rangle\langle\langle
c_{i\alpha\sigma}\mid c_{j\beta\sigma}^\dag\rangle\rangle]
 \nonumber\\
&&+U_1\langle n_{il\sigma}\rangle\langle\langle n_{i\alpha\bar{\sigma}}
c_{i\alpha\sigma}\mid c_{j\beta\sigma}^\dag\rangle\rangle,
 \\
(\omega +\mu_\alpha &-&U'\sum_{s\neq\alpha}\langle
n_{is}\rangle+J_z\sum_{s\neq\alpha}\langle
n_{is\sigma}\rangle)\langle\langle
n_{il\bar{\sigma}}c_{i\alpha\sigma}\mid
c_{j\beta\sigma}^\dag\rangle\rangle =\langle
n_{il\bar{\sigma}}\rangle\delta_{ij}\delta_{\alpha\beta}
-\sum_{b\in Fe,m}T_{ib\alpha m}\langle n_{il\bar{\sigma}}\rangle\langle\langle
c_{bm\sigma}\mid c_{j\beta\sigma}^\dag\rangle\rangle
 \nonumber\\
&&+U_1\langle n_{il\bar{\sigma}}\rangle\langle\langle
n_{i\alpha\bar{\sigma}}c_{i\alpha\sigma}\mid
c_{j\beta\sigma}^\dag\rangle\rangle -\sum_{b\in Cu}t_{ib}'(\langle
n_{il\bar{\sigma}}\rangle\langle\langle d_{b\sigma}\mid
c_{j\beta\sigma}^\dag\rangle\rangle +2\langle d_{b\bar{\sigma}}^\dag
c_{il\bar{\sigma}}\rangle\langle\langle c_{i\alpha\sigma}\mid
c_{j\beta\sigma}^\dag\rangle\rangle).
\end{eqnarray}
\end{widetext}

\bibliography{apssamp}

\end{document}